# The Multi-biophysical nature of Computation in brain neural networks

William Winlow and Andrew Simon Johnson

Dipartimento di Biologia, Università degli Studi di Napoli, Federico II, Via Cintia 26, 80126 Napoli, Italia
*Corresponding Author: bill.winlow@gmail.com or william.winlow@unina.it

**Abstract**

Comprehending the nature of action potentials is fundamental to our understanding of the functioning of nervous systems in general. The ionic mechanisms underlying action potentials in the squid giant axon were first described by Hodgkin and Huxley in 1952 and their findings have formed our orthodox view of how the physiological action potential functions. However, substantial evidence has now accumulated to show that the action potential is accompanied by a synchronized coupled soliton pressure pulse in the cell membrane, the action potential pulse (APPulse) which we have recently shown to have an essential function in computation. Here we explore the interactions between the soliton and the ionic mechanisms known to be associated with the action potential. Computational models of the action potential usually describe it as a binary event, but we have shown that it must be a quantum ternary event known as the computational action potential (CAP), whose temporal fixed point is the threshold of the soliton, rather than the rather plastic action potential peak used in other models to facilitate meaningful computation. We have demonstrated this type of frequency computation for the retina, in detail, and also provided an extensive analysis for computation for other brain neural networks. The CAP accompanies the APPulse and the Physiological action potential. Therefore, we conclude that nerve impulses appear to be an ensemble of three inseparable, interdependent, concurrent states: the physiological action potential, the APPulse and the CAP. However, while the physiological action potential is important in terms of neural connectivity, it is irrelevant to computational processes as this is always facilitated by the soliton part of the APPulse.

## Introduction

The nature of the nerve impulse has been the subject of detailed research since the time of Sir Charles Sherrington in the 1890s (Nobel Prize 1) (1) and Edgar Adrian (Nobel Prize 2)(2) who were jointly awarded the Nobel Prize in Physiology or Medicine in 1932. Sherrington discovered the nature of reflexes and Adrian provided evidence for the all-or-none law in nerves and muscles. However, it was not until 1952 that Hodgkin, Huxley (Nobel Prize 3) and Katz (3) revealed the ion-ic basis of the action potential itself in a brilliant series of experiments (Fig 1), supplemented by the work of Rall (1977) who provided the cable equation. Thus, the nerve impulse had been considered to be a largely electrochemical phenomenon and assumed the temporal and communicative aspects of nervous activity can be resolved by electrical theory. In other words, much has been done to understand the biophysical mechanisms underlying nerve action potentials, as a result of the excellent work by Hodgkin and Huxley (HH) and their predecessors.  However, it has become clear that action potentials are much more complex than originally suggested and mechanical, thermal and optical changes are known to be associated with them (e.g. Drukarch and Wilhelmus 2024(5) and see Table1) and should be incorporated into their description showing that action potentials have a multi-biophysical nature. Furthermore, action potentials have now been shown to have three interdependent functions of communication, modulation, and computation (6).

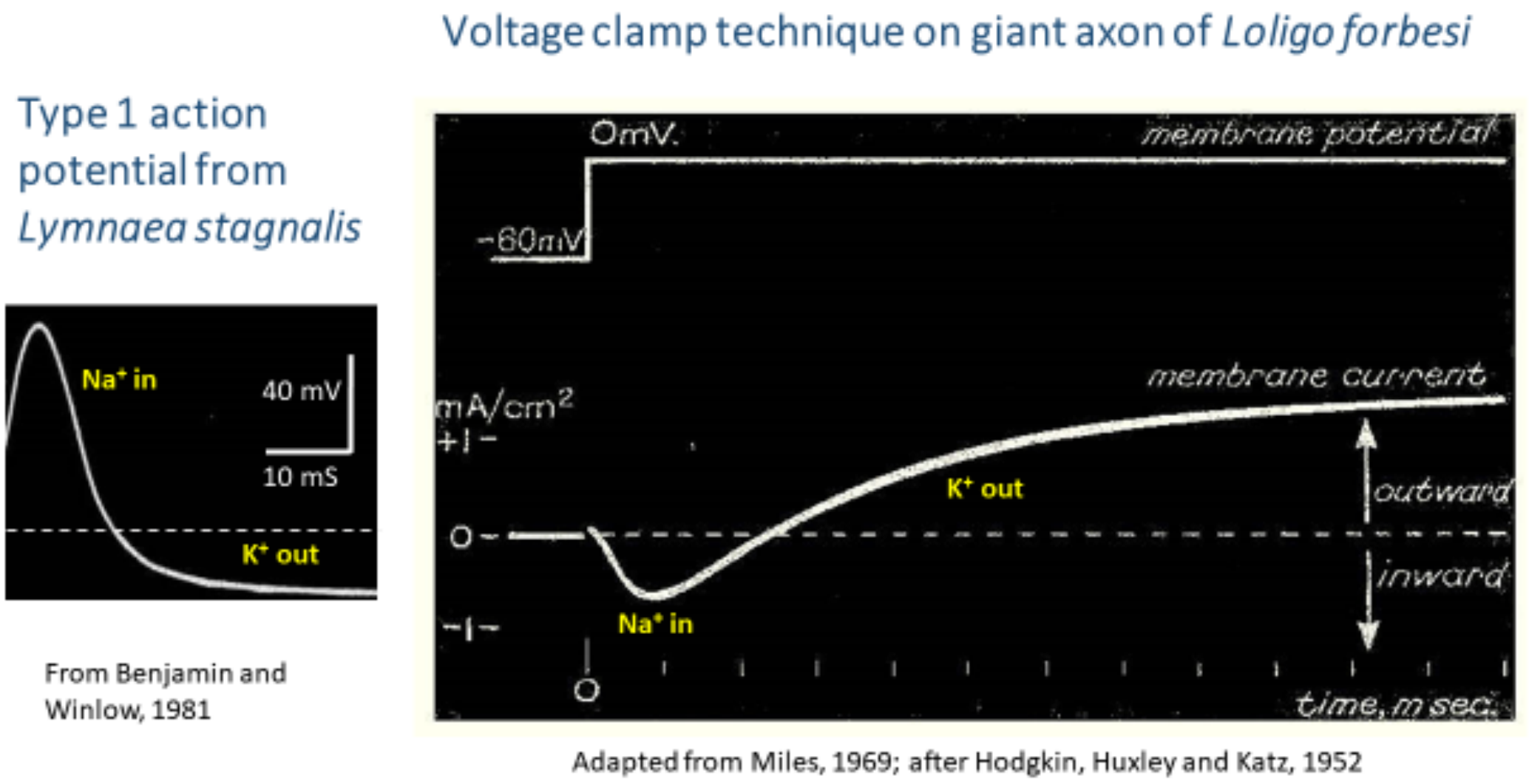

**Figure 1.** The action potential monitored in a snail neuron at left, using a DC preamplifier and a glass microelectrode to measure voltage changes, and using the voltage clamp technique to measure current changes at right, as visualised by Hodgkin, Huxley and Katz in 1952 (from Miles 1981) in a squid giant axon.  In both cases the dashed line represents the threshold at which an action potential is triggered. Note the prolonged after-hyperpolarisation in the *Lymnaea* neuron during the refractory period.

## 1 Communication within nervous systems

***Neurons and networks.*** Neurons are diverse and have many shapes, sizes and functions. They may have evolved from secretory cells in the early metazoa. We can envisage that as animal size increased the action potential evolved to control secretions at a distance (7; 8) although many local circuit neurons in both vertebrates and invertebrates are spikeless (9, 10, 11, 12, 13). The discovery of the nature of the action potential was critical to the development of modern neurophysiology but has been modelled as a binary event in computational brain networks. We have shown this to be an inappropriate premise for computation, both in AI and within the nervous system, and we will expand on the idea of phase ternary computation below.

***The action potential pulse (APPulse) is due to solitons accompanying action potentials***

In both axons (13,14,15), cardiac muscle (16), and most probably skeletal muscle, assumptions of almost instantaneous activation of progressive ion channels to produce the Hodgkin Huxley Action Potential are based upon the belief that electrostatic charge can travel from one channel to the next at the speed of the action potential. However, empirical evidence from channel spacing, ionic radii and diffusion coefficients demonstrate this is not, by itself, the case (13,14, 16). Substantial evidence has now accrued to show that the HH action potential is accompanied by a coupled soliton pulse (Table 1), which appears to instigate channel opening as shown in Figure 2 (from Johnson and Winlow 2018)(14). Unfortunately, the original work of Tasaki et al (1982) (17,18) was largely ignored until the early 21$^{st}$ century (19). The APPulse will of course require energy to maintain its progress and we suggest that this is generated by the ATP pumps that also generate the membrane potential and provide the energy for the HH action potential.

**Table 1. The Action Potential Pulse (APPulse) -** There is now a large body of evidence showing that:

- A'soliton' mechanical pulse accompanies an action potential and is stable propagating at constant velocity (17,18; El Hady et al., 2015; 19)
- Electrically recorded action potentials are accompanied by optically detected movements of the action potential, which mimic the action potential (21; 22)
- Non-linear sound waves/pressure pulses in lipid monolayers can show the main characteristics of nerve impulses (23; 24; 25; 26
- Membrane oscillator theory suggests that ion currents associated with action potential cause vortex phenomena leading to pressure waves in the nerve cell membrane (27)
- Ion channel separation is too great to allow for ion channel interference from adjacent channels caused by ionic charge (28,14; 29; 30; 31; 49)
- Ion channels can be opened by mechanical stimulus (32; 33; 34;)
- There is deformation of the membrane by activation of ion channels (17,18 20)
- Entropy (thermodynamic) measurements do not follow the H&H action potential but do follow the APPulse. (35; 18; 36; 37).
- These findings are supported by detailed mathematical modelling and computational simulations (38;)
- Further discussion: (5, 39 40,41)

**Table 1 – Accumulating evidence for the soliton pressure pulse in neuronal membranes.**

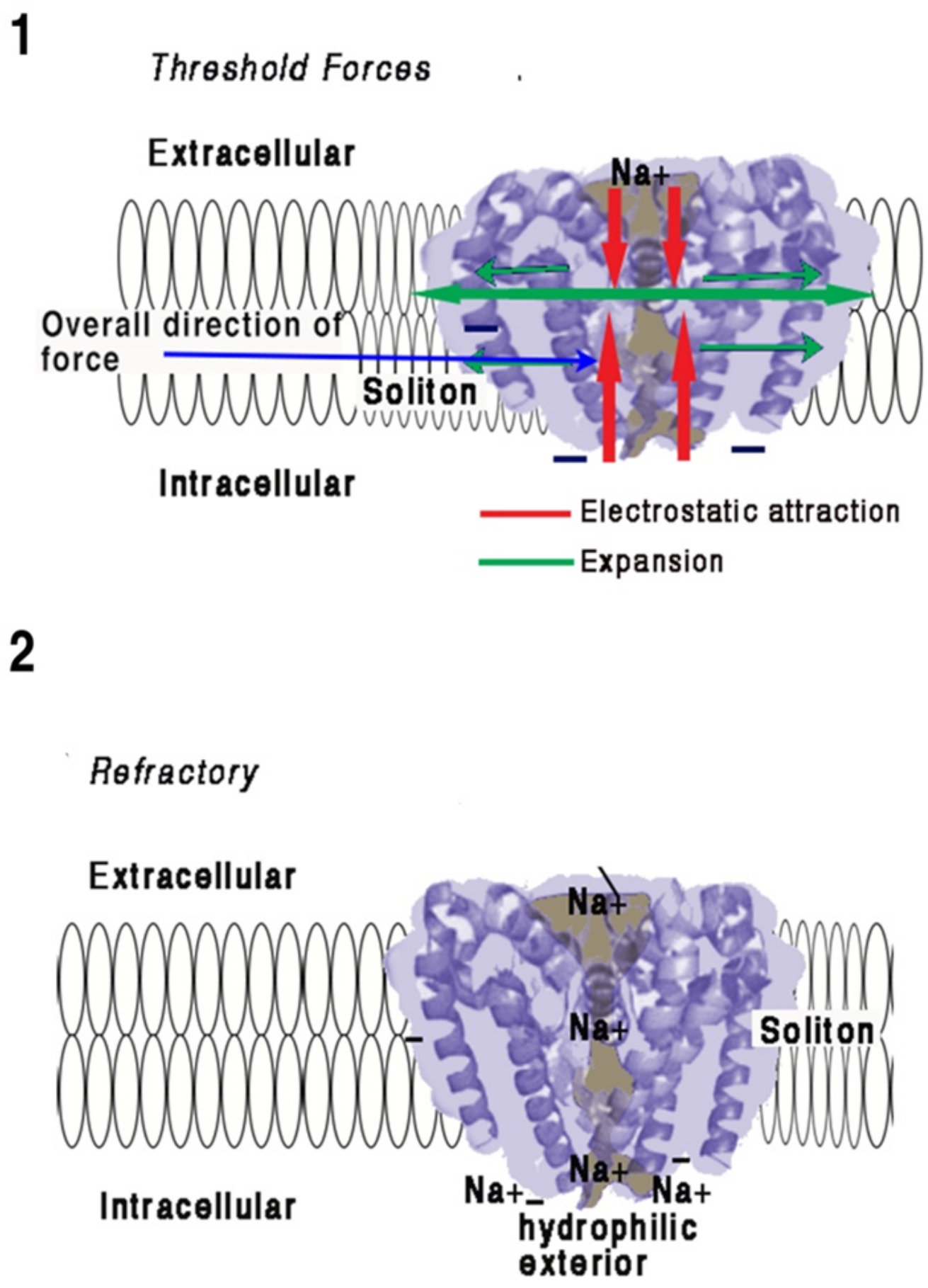

**Figure 2** - **Instigation of channel opening by the APPulse.** 1. Pressure from the accompanying pressure wave of the action potential disturbs the ion channel electrostatic seal. Attracted electrostatically charged ions pass through the channel causing it to contract across the membrane. This in turn puts energy back into the pressure wave. 2. The ion channel becomes refractory when enough Na+ ions pass through to produce electrostatic equilibrium. Partially reconstructed from Johnson and Winlow [28] and McCusker et al [42] , used under Creative Commons BY-NC-SA 3.0

## 2 Modulation

### *Chemical and electrical transmission*

One of the major hurdles to overcome in our understanding of the nervous system is that of memory storage after learning. Synapses provide latency changes through the actions of neurotransmitters resulting in an unpredictable timing for computation (40, 41). Although, electrical synapses have reduced latency (40) compared with chemical synapses, both together produce a spectrum of latencies depending upon their exact construction and location.

### *Learning and memory storage*

Currently, the role of synaptic plasticity in learning and memory is under intense scrutiny by cognitive scientists (43) with the suggestion that memories are molecularly stored within neurons and that plastic synaptic changes in weighting occur only after learning has occurred and been stored in memory.  Support for this idea comes from work by on *Lymnaea stagnalis* (44; 45) and on *Aplysia* (46) where it has been demonstrated that long term memory is stored in cell bodies. Thus, plastic changes in synaptic weighting may be "*a means of regulating behavior……..only after learning has already occurred*" (43).  Such assumptions would not be counter to findings that increased levels of associative learning efficiency are correlated with increased weighting of glutamatergic synapses and down-regulated by increased weighting of GABAergic inputs to neurons in the mouse barrel cortex.  Major memory storage systems clearly reside within the brains of vertebrates and may be stored within reverberatory circuits (see below and Johnson and Winlow 2023) (41).

## 3 Neurocomputation

### *The Action potential peak is not suitable for computational modelling in the brain*

Action potentials are highly plastic phenomena and vary greatly in trajectory from one neuron to the next. The temporal positioning of the spike peak is very variable (Figure 3) and modifiable by synaptic inputs.  Consequently, it is inappropriately used in binary computational models of neuronal activity. Here we demonstrate that only the action potential threshold taken from the beginning of the soliton has temporal constancy and therefore should be used in ternary computational models, whose phases are: resting potential, threshold and the time dependent refractory period, which is an analogue variable, formed by the recovery of the membrane, able to cancel all other action potentials in its path. Thus, the APPulse soliton threshold is the most appropriate temporal fixed point for computational modelling, not the action potential peak and not the loosely defined action potential threshold.

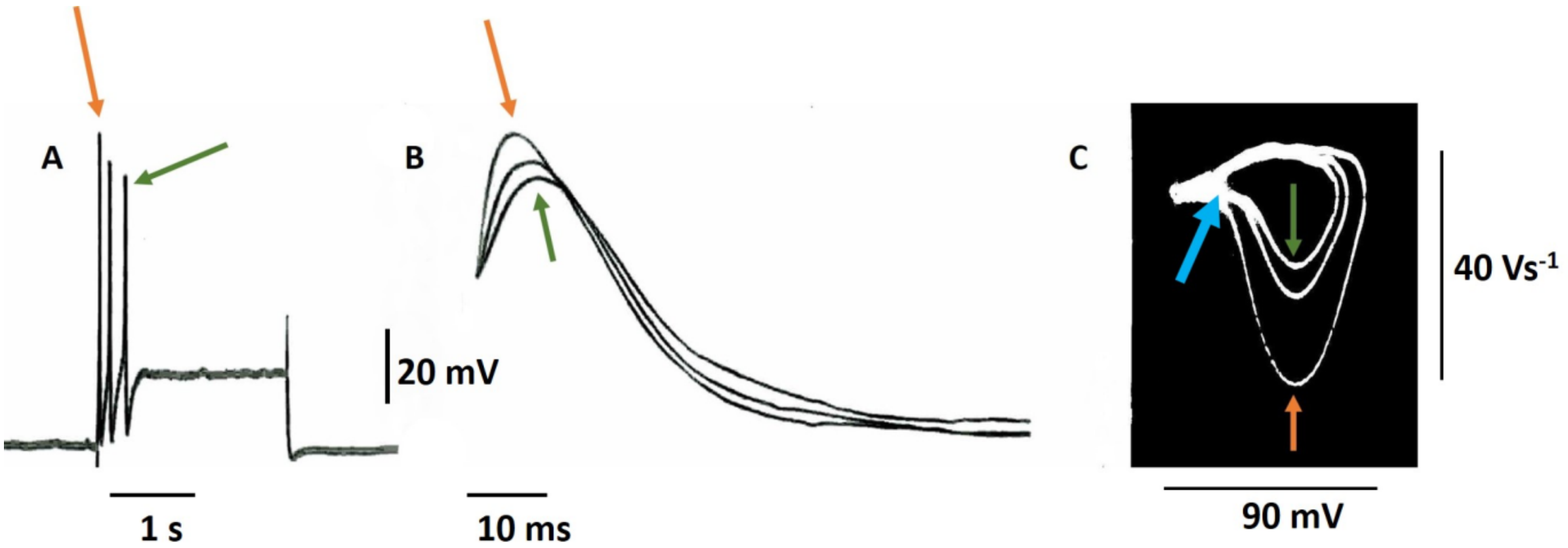

**Figure 3 – Plasticity of action potential shape and action potential peak** recorded from the soma of a fast adapting pedal I cluster neuron (for details see 15) in the intact brain of the mollusc *Lymnaea stagnalis* (L.). The cell was normally silent and activity was initiated by a 0.2nA current pulse of 3 s duration injected into the cell via a bridge balanced recording electrode. The same three spikes are represented in each case; a) on a slow time base, b) on a faster time base and c) as a phase plane portrait in which rate of change of voltage (dV/dt)is plotted against voltage itself and the inward depolarizing phase is displayed

downward maintaining the voltage clamp convention (see 11 for details of the phase plane technique). In each trace the peak of the first action potential is indicated by an orange arrow, the second action potential peak is unlabelled the third action potential peak is indicated by a green arrow. The three successive spike peaks clearly vary temporally from one another, but the threshold point of initiation remains constant as indicated in c) by the blue arrow in the phase plane portrait. From Winlow and Johnson, 2020 (40), licensed under Creative Commons BY-NC-SA 4.0)

***Neural transactions are performed by quantum phase ternary frequency computation.***

Historically the action potential has been viewed as a being a binary event and binary mathematics has been assumed to mediate computation in the brain. However, computation in a network may occur in a number of ways and a binary notation is not exclusive, logic may use other base forms and timings. Superficial calculations on timing, facility of computation and error demonstrate that only a few current models are applicable to vertebrate brains or those of advanced invertebrates and almost all can be immediately discounted (15, 40). In realistic neural networks and in the brain the only element of the action potential that is responsible for its live propagation is whatever mechanism causes the threshold at the leading edge. Thus the threshold alone is the initiator of the action potential (Fig 3) and its temporal marker. The timing of the spike is therefore directly related only to the threshold, which is a quantum event. Additionally, the threshold may be better defined temporally so that it is not a rising potential difference but a direct change over time. The rest of the action potential is only concerned with the refractory period and stabilisation to resting potential and is irrelevant to speed of transmission, although of course the refractory period can affect the frequency of transmission (40) and affect computation by backpropagation. In computational terms this means that the **computational action potential (CAP) is a phase ternary event** comprised of:

1. **Resting potential** of indeterminate length, actively maintained by membrane pumps
2. **Quantum threshold**– a temporal fixed point unlike the plastic action potential peak or the action potential threshold. Above this threshold a quantum of information, powered by the ATP pumps, passes through the neuron to the next synapse, with the HH spike arriving rather later as part of the refractory process.
3. **Refractory phase** when no new action potentials can occur and is an analogue variable, unlike the other two phases. It is able to reroute action potentials along different pathways at bifurcations ( see Johnson and Winlow 2025)(40).

The threshold is produced by the opening of sodium gates and is the rate limiter of a neuron (Figure 2). However no spike is required for computation, which can therefore occur even in spikeless neurons. The refractory period is only relevant when action potentials collide in phase-ternary computation (PTC) which is accurate to microseconds, i.e. it is both fast and efficient (40). Elsewhere, we have described how this may occur in the vertebrate retina (12,41). Furthermore, where electrical synapses occur the APPulse will pass unhindered from one neuron to the next. However, at chemical synapse a mechanical contact will be required for the pulse to pass. Glial cells such as astrocytes are perfectly positioned to provide such mechanical continuity (41) and doubtless there will be equivalent glial cells in invertebrates.

***Neurocomputation underlying perception and sentience in neocortex***

As demonstrated above, the physiological action potential is not sufficiently precise to use in computation, but the CAP is much more precise as it does not rely on the action potential peak but on the precise timing of the opening of the ion channels that provides entropy for the threshold of the soliton: the quantum threshold. It should be noted that the CAP is not binary, but a compound digital ternary object with an analogue third phase. The electro-physiologically recorded action potential has three phases: i) the resting potential, ii) the spike, whose peak is frequency dependent and highly plastic (Figure 3), iii) the refractory period – a time dependent analogue variable.
From what we have set out above it should be clear that the basis for computation in the brain is the quantum threshold of the 'soliton', which accompanies the ion changes of the action potential, and the refractory membrane at convergences. Knowing that we have been able to provide a logical explanation from the action potential to a neuronal model of the coding and computation of the retina (12). We have shown elsewhere that it automatically redacts error (13; 14, 41). This is particularly important in the parallel connections of the neocortex where memory is encoded in reverberatory loops of CAPs (47,41). However, unlike Turing based conventional computers and artificial intelligence (AI) there is no functioning clock in the nervous system and computation occurs by relative timing of the concurrent frequencies of quantum pulses. We showed in the retina timing from one point to depends upon transmission speed and the length of the neuron producing fixed latencies, which are both variable from one neuron to another. In this context therefore the quantum begins at the threshold of the soliton pulse and ends with the refractory period.

We have explained elsewhere how the visual cortex operates by quantum phase processing (47,41). We know vision is both robust and error free with the eye able to adapt during macular degeneration and this physiology is best explained using the CAP model (12,14,41). This is most likely to be true of hearing and other sensory systems which have similar connectivity, converging neuronal structures and and appropriate cortical neural network. The cortex of the brain structures resemble a small world networks, such as those described elsewhere (48), where connectivity is facilitated through multiple connections and each connection is temporally proximal to all other connections such that timing of computation is performed rapidly. In such a system parallel frequencies of APPulses are able to collide into definable patterns creating distinct object representations (12). Elsewhere in the eye and auditory systems, we have shown how many sensory cells are mean sampled to single neurons and that convergences of neurons are common. We have also demonstrated, using the threshold and refractory period of a quantum phase pulse, that action potentials diffract across a randomly formed neural network, such as the cortex, due to the annulment of parallel collisions in phase ternary computation (PTC). Thus, PTC applied to neuron convergences computes to collective mean sampled frequency and is the only mathematical solution within the constraints of brain neural networks (BNN) (12). In the retina and other sensory areas, we have discussed how this information is initially coded and then understood in terms of network abstracts within the lateral geniculate nucleus (LGN) and visual cortex. By defined neural patterning within a neural network by adjusting frequencies we have been able to theorise that abstraction occurs similarly to a contextual network where objects and thoughts are contextually compared, and memory stored geographically in the network: this idea will be familiar to network computer scientist and observers of brain activity. (41) The output of frequencies from the visual cortex represents information amounting to abstract representations of objects in increasing detail. The evidence of nerve tracts, from the LGN, can be extracted from this and we propose that these loops provide time synchronisation to the neocortex for the most basic representations. The full image is therefore combined in the neocortex with other sensory modalities, so that it receives concurrent information about the object from the eye, and all abstracts that make up the object. Spatial patterns in the visual cortex are formed from individual patterns illuminating the retina and memory is encoded by reverberatory loops of computational actions potentials (CAPs). We believe that a similar process of PTC may take place in the cochlea and associated ganglia, as well as ascending information from the spinal cord, and that this function should be considered universal where convergences of neurons occur.

Summary

Our findings are summarised in Figure 4

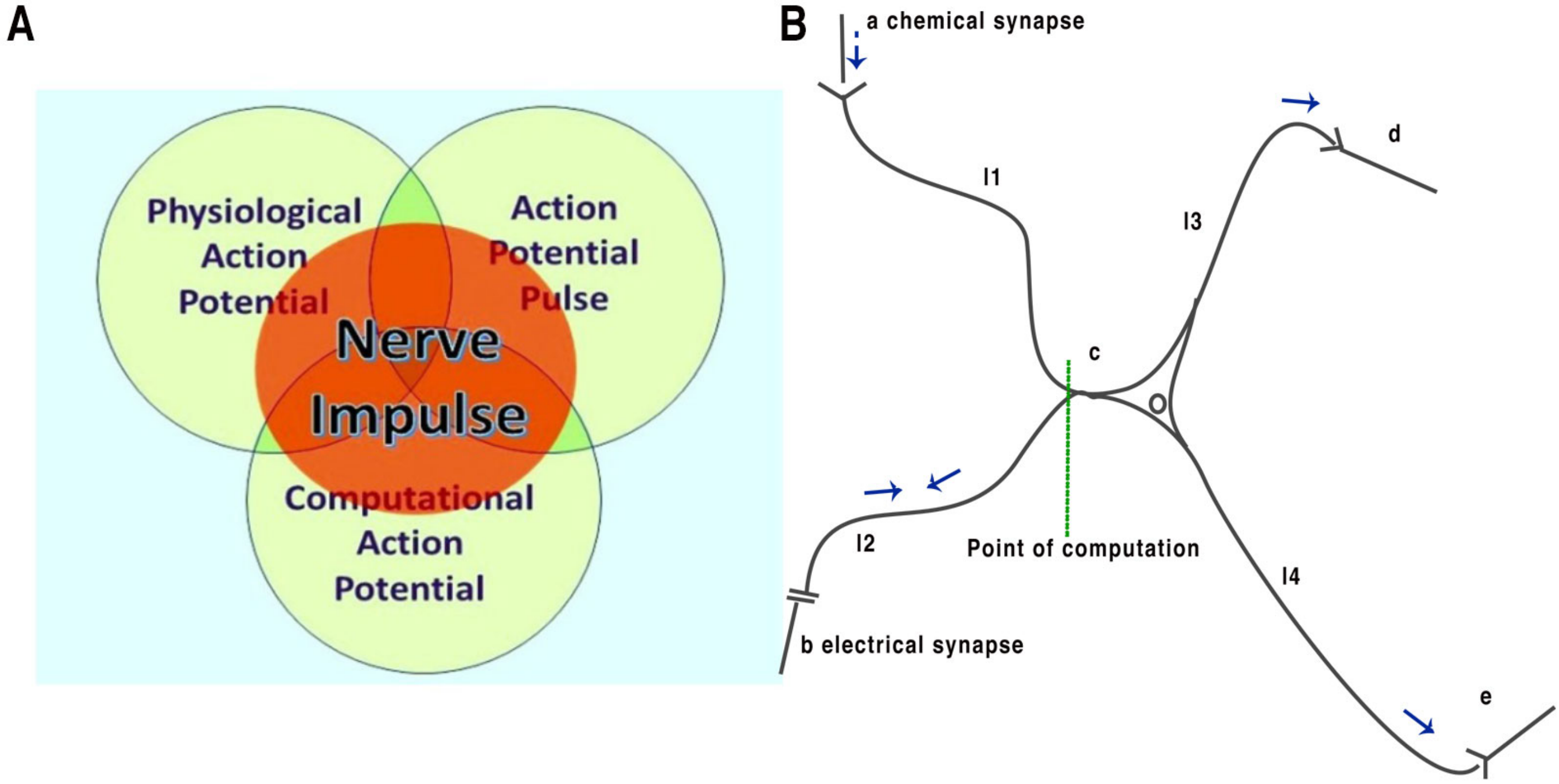

**Figure 4** A)The nerve impulse appears to be an ensemble of three inseparable components: the physiological action potential, as described by Hodgkin and Huxley (1952), the action potential pulse (Johnson and Winlow, 2018) and the computational action potential (From Winlow and Johnson 2021, with permission). B) illustrates computation over a single neuron c showing inputs from neurons a, b and outputs to d and e with a single point of computation (green). APPulse frequencies from a and combine or redact according to the rules of the Computational Action potential (CAP) forming changed frequencies outputted to d and e l1, l2 l3 and l4 indicate different latencies formed from both the different transmission lengths of the axons and the latencies of the synapses. Synapse b - c is an electrical synapse and is thus bidirectional and able to communicate by backpropagation to b. l1, l2, l3 and l4 are exaggerated for illustration and may facilitate spikeless computation (see APPulse) where only the threshold of the soliton becomes apparent.

**Conclusions**

Timing in the brain must account for the precision and activity of the neural networks and the ability to react to fast changes in stimulus and encompass learning. This functionality cannot be replicated by the action potential alone as it does not have sufficient precision or timing necessary. Furthermore, no method of achieving this has been discovered within the brain. However, the CAP formed from the APPulse soliton threshold acting as a frequency modulated computer can be shown building on the work of HH to have this ability and to diffract and form the necessary memory and activity observed in the brain in concert with the refractory period observed in the HH action potential. In other words, the HH action potential cannot act alone, and computation must be facilitated by the leading edge of the soliton.